\documentclass[12pt]{iopart}
\usepackage{graphicx}

\begin{document}

\title{Quantitative measures of entanglement in pair coherent states}

\author{G. S. Agarwal$^1\footnote{On leave of absence from Physical Research
Laboratory, Navrangpura, Ahmedabad, India}$ and Asoka Biswas$^2$}
\address{$^1$Department of Physics, Oklahoma state University,
Stillwater, OK - 74078, USA}
\address{$^2$Physical Research Laboratory,
Navrangpura, Ahmedabad - 380 009, India}
\eads{\mailto{agirish@okstate.edu}, \mailto{asoka@prl.ernet.in}}
\date{\today}

\begin{abstract}
The pair coherent states for a two-mode radiation field are known
to belong to a family of states with non-Gaussian wave function.
The nature of quantum entanglement between the two modes and some
features of non-classicality are studied for such states. The
existing criteria for inseparability are examined in the context
of pair coherent states.
\end{abstract}

\pacs{03.67.Mn, 42.50.Dv}

\maketitle

\section{Introduction}

Entanglement in continuous variables has been of great interest
since the celebrated paper of Einstein, Podolsky and Rosen (EPR)
\cite{einstein} who constructed a two-particle state which was
strongly entangled both in position and momentum spaces. The EPR
state has been the subject of many discussions on the nonlocality
of quantum mechanics. It turns out that the EPR state can be
physically realized in a high gain parametric amplifier. This
opened up the possibility of a variety of new experiments
\cite{ou,howell,kimble,kim} using entanglement in continuous
variables. The Wigner function for such states is Gaussian in
position and momentum variables \cite{foot1}. The Gaussian states
are very special in the sense that the information on the higher
order correlations can be extracted from second order
correlations. The criteria for entanglement these states has been
formulated in terms of second order correlations between position
and momentum variables \cite{duan,simon}. Mancini {\it et al.\/}
\cite{mancini} derived an equivalent set of criteria. A great
advantage of these inequalities is that the transpose criteria has
been translated into something which is directly measurable. In
this paper we focus our attention on the entangled character of a
family of non-Gaussian states, viz., the pair coherent states.
These states are entangled since the expression for pair coherent
states is already in Schmidt form. We examine its entanglement
character in terms of the Peres-Horodecki criteria. We calculate
explicitly the eigenvalues of the partial transpose of the density
matrix and show that some of these are negative. We also present
results on the correlation entropy and linear entropy.

To setup an experiment to quantitatively measure entanglement, one
can think of quasi-probability distribution functions, viz. the
Glauber-Sudarshan P-function, Q-function, and the Wigner function.
We study the relationship between the non-classicality the
P-function of the state and entanglement. It is known that if
P-function of the state is well-behaved, then the state is
separable. We study the P-function for the pair coherent state to
show the entangled nature of the state. we also check the
entanglement in pair coherent state using the inequalities for
second order moments.

The organization of this paper is as follows. In Sec. II, we
introduce the family of bipartite non-Gaussian states of radiation
field. We describe their properties briefly. In Sec. III, we
investigate the inseparability of the pair coherent states in
light of Peres-Horodecki criterion and von Neumann entropies.
Later we discuss the relation between entanglement and
non-classicality of P-function. Finally we study the existing
separability inequality to detect entanglement in the pair
coherent state.

\section{Pair coherent state: An entangled non-Gaussian state}
The simplest examples of non-Gaussian states of the field are,
say, the single photon states. Other examples could be states
generated by excitations on a Gaussian state
\cite{tara_thermal,gsapuri}. The state which has been extensively
studied for its nonclassical properties and violation of Bell
 inequalities \cite{gilchrist,tara} is the pair coherent state \cite{gsa_pc}.
A pair coherent state $|\zeta, q\rangle$ is the state of a
two-mode radiation field \cite{gsa_pc} with the following
properties:
\numparts
\begin{eqnarray}
a b|\zeta, q\rangle =\zeta |\zeta,q\rangle\;,\\
(a^\dagger a-b^\dagger b)|\zeta,q\rangle =q|\zeta,q\rangle\;,
\end{eqnarray}
\endnumparts
where $a$ and $b$ are the annihilation operators associated with
two modes, $\zeta$ is a complex number, and $q$ is the degeneracy
parameter. The pair coherent state for $q=0$ (corresponding to
equal photon number in both the modes) is given by
\begin{equation}
|\zeta,0\rangle=N_0\sum_{n=0}^\infty
\frac{\zeta^n}{n!}|n,n\rangle\;, \label{pcs}
\end{equation}
where $N_0=1/\sqrt{I_0(2|\zeta|)}$ and $I_0(2|\zeta|)$ is the
modified Bessel function of order zero. The coordinate space wave
function is given by
\begin{eqnarray}
\langle x_a,x_b|\zeta,0\rangle
&=&N_0\sum_{n=0}^\infty\frac{\zeta^n}{n!}\langle
x_a|n\rangle\langle x_b|n\rangle\nonumber\\
&=&N_0\sum_{n=0}^\infty\frac{\zeta^n}{n!}\frac{1}{\sqrt{\pi}}\frac{H_n(x_a)H_n(x_b)}{2^nn!}\exp\left[-\frac{x_a^2+x_b^2}{2}\right]\;,
\label{space_pcs}
\end{eqnarray}
where $\langle x_a|n\rangle$ is a harmonic oscillator wave
function given in terms of the Hermite polynomial as
\begin{equation}
\label{xa_n}\langle
x_a|n\rangle=\left(2^nn!\sqrt{\pi}\right)^{-1/2}H_n(x_a)e^{-x_a^2/2}\;.
\end{equation}
\begin{figure}
\centerline{\scalebox{0.5}{\includegraphics{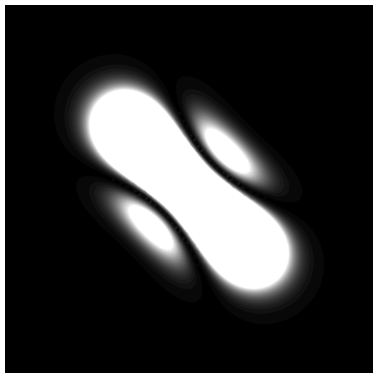}}}
\caption{\label{fig0}Contour plot of the quadrature distribution
$P(x_a,x_b)$ for the pair coherent state for $\zeta=-1$.}
\end{figure}
It is clear from the expression (\ref{space_pcs}) that the wave
function of the pair coherent state is non-Gaussian. We have shown
the quadrature distribution $P(x_a,x_b)=|\langle
x_a,x_b|\zeta,0\rangle |^2$ of the state (\ref{space_pcs}) in
Figure \ref{fig0}. The distribution reflects the entanglement
present in the state. Note that the pair coherent state can be
obtained by projecting the two-mode coherent state
\begin{equation}
|\alpha,\beta\rangle=e^{-(|\alpha|^2+|\beta|^2)/2}\sum_{n,m=0}^\infty
\frac{\alpha^n\beta^m}{\sqrt{n!m!}}|n,m\rangle
\label{two_mode}
\end{equation}
onto a space with a fixed difference of the number photons in two
modes. The well known squeezed vacuum state with Gaussian wave
function is given by
\begin{equation}
|\zeta\rangle_{\mathrm{TP}}=\sqrt{1-|\zeta|^2}\sum_{n=0}^\infty\zeta^n|n,n\rangle\;.
\label{sqz_vac}
\end{equation}
Note that the expansion coefficients are different in (\ref{pcs})
where the coefficients decrease quickly with increase in $n!$.

\section{Inseparability of the Pair Coherent State}
In this section we quantitatively study the entanglement in pair
coherent state. Note that the state (\ref{pcs}) has an obvious
form of Schmidt decomposition. This reflects the fact that this
state is an entangled state. In the next subsections, we examine
the other criteria to give an estimate of its entanglement.

\subsection{Peres-Horodecki inseparability criteria} The
Peres-Horodecki inseparability criterion \cite{peres} is known to
be necessary and sufficient for the $(2\times 2)$ and $(2\times
3)$ dimensional states, but to be only sufficient for any higher
dimensional states. This criterion states that if the partial
transpose of a bipartite density matrix has at least one negative
eigenvalue, then the state becomes inseparable. The density matrix
$\rho$ corresponding to the state $|\zeta,0\rangle$ (which is a
infinite dimensional state) can be written as
\begin{equation}
\rho=\left(\sum_{n=0}^\infty C_{nn}|n,n\rangle\right)\left(\sum_{m=0}^\infty C_{mm}^*\langle m,m|\right)\;,
\end{equation}
where $C_{mm}=N_0\frac{\zeta^m}{m!}$. Hence the partial transpose
of $\rho$ is given by
\begin{equation}
\rho_{\textrm{PT}}=\sum_{n,m=0}^\infty
C_{nn}C_{mm}^*|n,m\rangle\langle m,n|\;. \label{rho_pt}
\end{equation}
One can now calculate the eigenvalues and eigenfunctions of the
matrix $\rho_{\textrm{PT}}$ as follows: Let us start with the
following set of two Hermitian conjugate terms for $n\neq m$ in
the above equation
\begin{equation}
C_{nn}C_{mm}^*|n,m\rangle\langle
m,n|+C_{mm}C_{nn}^*|m,n\rangle\langle n,m|\;.
\end{equation}
Diagonalizing the above block of the matrix $\rho_{PT}$ we find
the following eigenvalues:
\begin{eqnarray}
\lambda_{nn}&=&\frac{1}{I_0(2|\zeta|)}\frac{|\zeta|^{2n}}{(n!)^2}\;,\;\;\;\forall n\nonumber\\
\lambda_{nm}^\pm
&=&\pm\frac{1}{I_0(2|\zeta|)}\frac{|\zeta|^{n+m}}{n!m!}\;,\;\;\forall
n\neq m\;. \label{eigenvalues}
\end{eqnarray}
and the corresponding eigenfunctions $|n,n\rangle$ and
$(|n,m\rangle\pm e^{-i\theta}|m,n\rangle)/\sqrt{2}$, where
$\theta$ is the relative phase of the amplitudes $C_{nn}$ and
$C_{mm}$ and is defined by
$e^{i\theta}=C_{nn}C_{mm}^*/|C_{nn}||C_{mm}|$. Clearly the matrix
$\rho_{\textrm{PT}}$ has several negative eigenvalues. Hence
according to the Peres-Horodecki criterion, the pair coherent
state is an inseparable state\footnote{The eigenvalues of the
partial transpose of the density matrix for the squeezed vacuum
state are also given by (\ref{eigenvalues}) with $n!$ and $m!$
replaced by unity.}. Note that if the phase of the parameter
$\zeta$ is random, then the state becomes separable, as then terms
corresponding to different values of $n$ and $m$ drop out of the
double summation in (\ref{rho_pt}).

\begin{figure}
\centerline{\scalebox{0.4}{\includegraphics{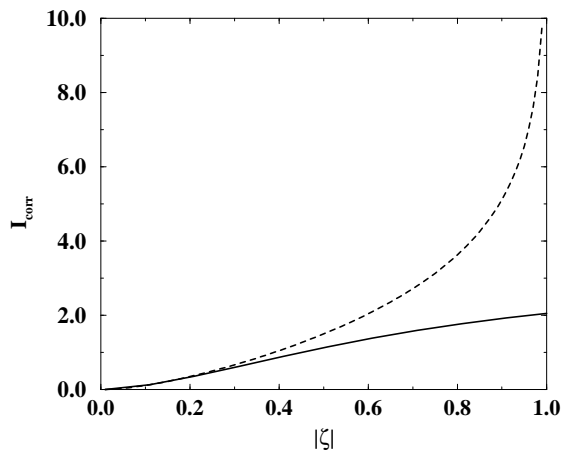}}}
\caption{\label{fig2}Variation of the correlation entropy of the
pair coherent state (solid line) and the squeezed state (dashed
line) with $|\zeta|$. For large values of $|\zeta|$ the
correlation entropy varies linearly with $|\zeta|$.}
\end{figure}

\subsection{Correlation entropy}
Correlation entropy of a
bipartite system consisting of subsystems $a$ and $b$ is given by
\cite{huanggsa}
\begin{equation}
I_{\mathrm{corr}}=S_a+S_b-S_{ab}\;,
\label{icorr}
\end{equation}
where $S_k$ is the von Neumann entropy of the system $k$. If $a$ and $b$ are uncorrelated
(separable), then $I_{\mathrm{corr}}$ vanishes.
Now for
any bipartite pure state, $S_{ab}$ is zero. We have calculated $S_{a,b}$ for the
pair coherent state as
\begin{equation}
S_a=S_b=-\sum_{n=0}^\infty\frac{|\zeta|^{2n}}{I_0(2|\zeta|)n!^2}\textrm{log}_2\left(\frac{|\zeta|^{2n}}{I_0(2|\zeta|)n!^2}\right)
\;. \label{icorr_pcs}
\end{equation}
We plot the correlation entropy (\ref{icorr}) for the pair
coherent state with $|\zeta|$ in Figure \ref{fig2}. The
correlation entropy for the pair coherent state remains non-zero
for all values of $|\zeta|$ which signifies that the state is
inseparable (entangled) for all $|\zeta|$. For smaller values of
$|\zeta|$, the entropy increases slowly; but at larger values of
$|\zeta|$, it saturates. Note that the squeezed vacuum state
(\ref{sqz_vac}) is a very special kind of Gaussian non-classical
state, entanglement properties of which have been much studied in
literature \cite{howell}. For a better understanding of the
inseparability of the pair coherent state, we have compared the
correlation entropies of the pair coherent state and of the
squeezed vacuum state in Figure \ref{fig2}. Clearly the
correlation entropy of the squeezed vacuum state grows much faster
than that of the pair coherent state. This is because, as
$|\zeta|$ approaches to unity, the squeezed vacuum state becomes
much more incoherent than the pair coherent state. At $|\zeta|=1$
the squeezed vacuum state becomes completely random, as all the
possible states of the either mode then become equally probable.
It is well-known that the correlation entropy of a system with all
basis states equally probable is $2\log_2N$, where $N$ is the
number of possible basis states of the system. For squeezed vacuum
state, as $N\rightarrow\infty$, the correlation entropy diverges
for $|\zeta|=1$. On the other hand, in case of pair coherent
state, the picture is different at $|\zeta|=1$. In this case the
states with lower occupation number become more probable than the
states with higher occupation number. Thus the correlation entropy
for the pair coherent state remains less than that of the squeezed
vacuum state for $|\zeta|$ approaching unity, suggesting that the
pair coherent state remains much more coherent than the squeezed
vacuum.

\begin{figure}
\centerline{\scalebox{0.4}{\includegraphics{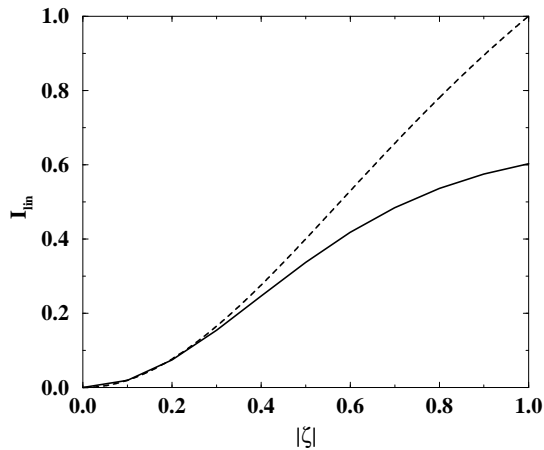}}}
\caption{\label{fig1a}Variation of the linear entropy
$I_{\mathrm{lin}}$ for pair coherent state (solid line) and
squeezed vacuum state (dashed line) with $|\zeta|$.}
\end{figure}

\begin{figure}
\centerline{\scalebox{0.45}{\includegraphics{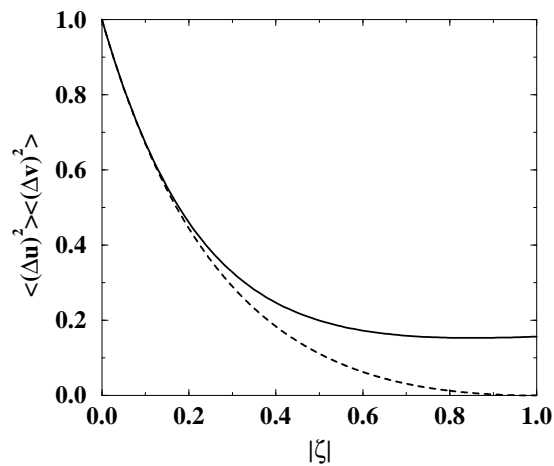}}}
\caption{\label{fig1} Variation of product of the variances of the
joint position and momentum of two subsystems in a pair coherent
state (solid line) and in a two-mode squeezed vacuum state (dashed
line) with the squeezing parameter $|\zeta|$, for $\phi=\pi$ and
$m=1$. The product remains less than unity for all non-zero
$|\zeta|$.}
\end{figure}

\begin{figure}
\centerline{\scalebox{0.4}{\includegraphics{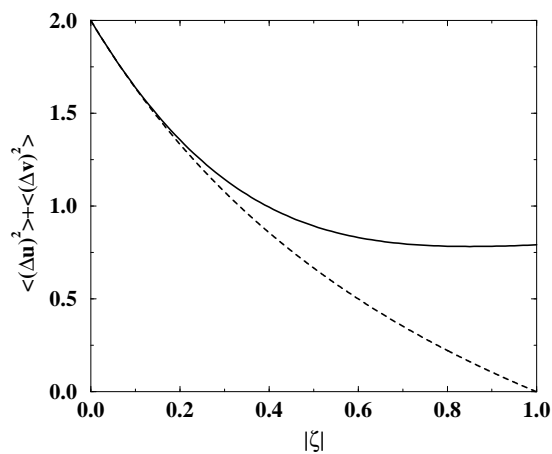}}}
\caption{\label{fig3}Variation of total variance in the EPR-like
variables $u$ and $v$ with $|\zeta|$ for pair coherent state
(solid line) and squeezed vacuum state (dashed line) for $m=1$. We
have chosen $\phi=\pi$.}
\end{figure}

\subsection{Linear entropy}
We further calculate the linear entropy of the pair coherent
state, which is given by
$I_{\mathrm{lin}}=1-\textrm{Tr}(\rho_k^2)$, where $\rho_k$ is the
reduced density matrix of the subsystem $k$. For a pure state
density matrix $\rho$, $I_{\mathrm{lin}}$ vanishes as
$\textrm{Tr}(\rho^2)=1$. But for an entangled state, $\rho_k$ does
not have the form of a pure state density matrix. Thus, any
non-zero $I_{\mathrm{lin}}$ provides signature of entanglement
present in the state. Note further that the linear entropy is
closely related to the entanglement measure in terms of Schmidt
number \cite{law}. For a pair coherent state, the linear entropy
is given by
\begin{equation}
I_{\mathrm{lin}}=1-\frac{1}{I_0(2|\zeta|)^2}\sum_{n=0}^\infty\frac{|\zeta|^{4n}}{n!^4}\;.
\end{equation}
We show the variation of the quantity $I_{\mathrm{lin}}$ with
$|\zeta|$ in Figure \ref{fig1a}. Clearly for any non-zero
$|\zeta|$, $I_{\mathrm{lin}}$ becomes non-zero and positive
implying entanglement in the pair coherent state. For $|\zeta|\sim
0.4$, $I_{\mathrm{lin}}\sim 0.25$, i.e., $\textrm{Tr}(\rho_k^2)$
is close to unity. This implies that the state represented by
$\rho_k$ is more like a pure state than a mixed state. On the
other hand, for $|\zeta|= 1$, $I_{\mathrm{lin}}\sim 0.6$, which
means that $\rho_k$ represents more like a mixed state than a pure
state. Thus the state (\ref{pcs}) is more entangled. Further we
have plotted the linear entropy of the squeezed vacuum state
(\ref{sqz_vac}) in Figure \ref{fig1a}. Clearly for small $|\zeta|$
the linear entropy of both the states show similar behavior. But
at larger values of $|\zeta|$, linear entropy of squeezed vacuum
state is more than the pair coherent state, i.e., the degree of
mixedness of the pair coherent state remains lower than the
squeezed vacuum state for larger values of $|\zeta|$.

\subsection{Entanglement and non-classicality of the $P$ function}
So far we have investigated the entanglement in pair coherent
state in terms of Peres-Horodecki inseparability criterion and
various entropies. However all these criteria are not possible to
verify in experiments. We will now the study the non-classicality
of P-function for the pair-coherent state as P-function of a
density matrix can be measured in experiments. For bipartite
systems, the two-mode density matrix can be written in terms of
the diagonal coherent state representation as
\begin{equation}
\rho=\int\int d^2\alpha d^2\beta
P(\alpha,\alpha^*;\beta,\beta^*)|\alpha,\beta\rangle\langle\alpha,\beta|\;.
\end{equation} It is known that if the $P$-function has
non-classical character, then the state is entangled. We examine
the inseparability of the pair-coherent state from the point of
view of the non-classicality of the $P$-function. The
Glauber-Sudarshan P-distribution function gives a
quasi-probability distribution in phase space, which can assume
negative and singular values for non-classical fields. There is
another distribution function called Q-function which is related
to the P-function by
\begin{eqnarray}
Q(\alpha,\alpha^*;\beta,\beta^*)&=&\frac{1}{\pi^2}\langle\alpha
,\beta|\rho|\alpha,\beta\rangle \nonumber\\
&=&\frac{1}{\pi^2}\int
P(\gamma,\gamma^{*};\delta,\delta^*)e^{-|\gamma-\alpha|^2}e^{-|\delta-\beta|^2}d^2\gamma
d^2\delta\;, \label{relation}
\end{eqnarray}
which is always positive. Note that if the function
$P(\gamma;\delta)$ were like classical probability distribution,
then $Q(\alpha;\beta)>0\;\forall \;\alpha,\beta$. However if $Q$
is zero, then $P$ must become at least negative (referring to
nonclassicality of the state) in some parts. Hence the exact
zeroes of the Q-function are also a signature for the
non-classicality of the field. In order to see the
non-classicality of the pair coherent state, we examine the
structure of the $Q$ function. The $Q$ function for the pair
coherent state can be calculated as
\begin{equation}
Q(\alpha,\alpha^*;\beta,\beta^*)=\frac{1}{\pi^2I_0(2|\zeta|)}e^{-(|\alpha|^2+|\beta|^2)}|I_0(2\sqrt{\zeta\alpha^*\beta^*})|^2\;.
\label{Q_pcs}
\end{equation}
This function has zeroes only if $\alpha$ and $\beta$ are out of
phase for real positive $\zeta$, and if
$2\sqrt{\zeta|\alpha||\beta|}=z_0$, where $z_0$ are the exact
zeroes of the Bessel function $J_0(z)$. The smallest few values of
$z_0$ are 2.4048, 5.52, 8.6537, 11.7915, 14.9309 etc. Existence of
these zeroes proves that the pair-coherent state is a
non-classical state.

However it is worth mentioning that, the notion that, if the
P-function of a state is not well-behaved, then the state in
inseparable, is not always true. For example, for a two-mode
separable Fock state $|n,m\rangle$, the P-function is not at all
well-behaved. This motivates us to investigate some alternative
inseparability criteria, which can be verified in experiments. We
will show that existing inseparability inequalities are quite
useful in detecting entanglement in non-Gaussian state like
(\ref{pcs}) in experiments, albeit under certain condition.

\subsection{Separability inequalities}
Duan {\it et al.\/} \cite{duan} and Simon \cite{simon}
independently have derived the separability criterion of a
bipartite continuous-variable system in terms of the second-order
correlations. This criterion states that if a state is separable,
then the uncertainties in a pair of EPR-like operators $u$ and $v$
satisfy,
\begin{equation}
\label{duan_crit}M=\langle (\Delta u)^2\rangle+\langle (\Delta
v)^2\rangle\geq m^2+\frac{1}{m^2}\;,
\end{equation}
where \numparts
\begin{eqnarray}
u&=&|m|x_a+\frac{1}{m}x_b\;,\\
v&=&|m|p_a-\frac{1}{m}p_b\;,
\end{eqnarray}
\endnumparts
for any arbitrary nonzero real number $m$. Here
$x_k=(k+k^\dag)/\sqrt{2}$ and $p_k=(k-k^\dag)/i\sqrt{2}$ ($k=a,b$)
are the position and momentum operators for the subsystem $k$
satisfying the commutation relation $[x_k,p_{k'}]=i\delta_{kk'}$.
Conversely, violation of this criterion provides a sufficient
condition for inseparability of states, albeit with a lower bound
\begin{equation}
\left|m^2-\frac{1}{m^2}\right|\leq M<m^2+\frac{1}{m^2},
\label{duan1a}
\end{equation}
which for $m=\pm 1$, reads as
\begin{equation}
\label{duan1}0\leq M<2\;.
\end{equation}
For a bipartite Gaussian state, the criterion (\ref{duan_crit}) is
also sufficient for separability.

Equivalent necessary and sufficient condition for separability of
Gaussian states have been derived by Englert and Wodkiewicz
\cite{englert} using density operator formalism. They have shown
that the positivity of the partial transposition and
P-representability of the separable Gaussian states are closely
related. These criteria have been experimentally verified via the
interaction of linearly polarized field with cold atoms
\cite{josse}, in atomic ensembles \cite{polzik}, and with squeezed
light fields \cite{bowen,silberhorn}. Mancini {\it et al.\/}
\cite{mancini} have shown that separability of a state leads to
the following uncertainties in a pair of EPR-like variables
\begin{equation}
\label{mancini}M_x=\langle (\Delta u)^2\rangle\langle (\Delta
v)^2\rangle\geq 1\;,\;\;\;m=1\;,
\end{equation}
where $u=x_a+x_b$ and $v=p_a-p_b$. Violation of this inequality
provides a sufficient criterion of inseparability in Gaussian
states.

We will now discuss the validity of the criterion (\ref{mancini})
in case of a pair coherent state which is a non-Gaussian state. We
calculate the uncertainties $\langle(\Delta u)^2\rangle$ and
$\langle(\Delta v)^2\rangle$ (for $m=1$) for the pair coherent
state $|\zeta,0\rangle$. We find the averages $\langle
x_a+x_b\rangle=0$ and
\begin{eqnarray}
\langle(x_a+x_b)^2\rangle&=&\langle\zeta|(1+a^\dag a+b^\dag b+ab+a^\dag b^\dag)|\zeta\rangle\nonumber\\
&=&1+2|\zeta|\frac{I_1(2|\zeta|)}{I_0(2|\zeta|)}+2|\zeta|\cos\phi\;.
\label{variancex_pcs}
\end{eqnarray}
%where we have used the relations
%\begin{equation}
%\left.\begin{array}{ccc}
%x_k&=&\frac{k+k^\dag}{\sqrt{2}}\;,\\
%p_k&=&\frac{k-k^\dag}{\sqrt{2}i}\;,
%\end{array}\right\}\;\;k=a,b\;.
%\label{xkpk}
%\end{equation}
Here $I_1(2\zeta)$ is the modified Bessel function of order one
and $\zeta=|\zeta|e^{i\phi}$. Thus $\langle(\Delta
u)^2\rangle=\langle u^{2}\rangle-\langle
u\rangle^2=\langle(x_a+x_b)^2\rangle$. In a similar way, one can
calculate the variance $\langle(\Delta v)^2\rangle$ which is found
to be equal to $\langle(\Delta u)^2\rangle$. In Figure \ref{fig1},
we have shown the variation of product $\langle(\Delta
u)^2\rangle\langle(\Delta v)^2\rangle$ with $|\zeta|$. Clearly,
the inequality (\ref{mancini}) is violated for all $|\zeta|$. We
can thus infer that the pair coherent state is an inseparable
state.

Now we will discuss whether the criterion [Equation
(\ref{duan1a})] is applicable for the pair coherent state. The
total variance $M=\langle (\Delta u)^2\rangle+\langle(\Delta
v)^2\rangle$ can be calculated as
\begin{eqnarray}
M&=&\left(|m|^2+\frac{1}{m^2}\right)+2\left(|m|^2+\frac{1}{m^2}\right)|\zeta|\frac{I_1(2|\zeta|)}{I_0(2|\zeta|)}\nonumber\\
&+&4\frac{|m|}{m}|\zeta|\cos\phi\;.
\end{eqnarray}
We show the variation of the above quantity with $|\zeta|$ in
Figure \ref{fig3} for $\phi=\pi$. The figure shows that total
variance remains less than $|m|^2+\frac{1}{m^2}$ for all
$|\zeta|$. Thus the inequality (\ref{duan1a}) is satisfied for the
pair coherent state under the condition
\begin{equation}
(\mathrm{sign\;of}\;m)(\mathrm{sign\;of}\;\cos{\phi})<0\;.
\end{equation}
Thus the criterion (\ref{duan1a}) is sufficient for inseparability
of the pair-coherent state only if the above condition is
satisfied.

We have also compared the degree of violation of (\ref{mancini})
and (\ref{duan1a}) of the pair-coherent state with that of the
squeezed vacuum state (\ref{sqz_vac}) in Figure \ref{fig1} and
Figure \ref{fig3}. It shows that the criteria (\ref{mancini}) and
(\ref{duan1a}) are violated more by the squeezed vacuum state for
larger $|\zeta|$. Thus the degree of inseparability is more for
the squeezed vacuum state at larger $|\zeta|$. This is due to the
fact that the expansion coefficients in the pair coherent state
[see Equation (\ref{pcs})] decrease quickly with the increase of
$n!$.

\section{Conclusions}
In conclusions, we have studied the inseparability of a special
family of non-classical states, called the pair coherent states
which are non-Gaussian in nature. We confirmed the inseparability
of pair coherent states in the light of Peres-Horodecki criteria
and various entropies. We then demonstrated that the existing
inseparability criterion (\ref{duan1a}) based on second order
correlation, is applicable to these kind of non-Gaussian states
only under certain constraints.

\end{document}